\documentclass{article}

\usepackage[utf8]{inputenc}
\usepackage{amsmath,amssymb,amsfonts,amsthm,tensor}
\usepackage{mathrsfs}
\usepackage{bbold}
\usepackage{cite}
\usepackage{verbatim}
\usepackage{float}
\usepackage{color}
\usepackage[a4paper, total={6in, 8in}]{geometry}
\usepackage{appendix}
\usepackage{hyperref}
\usepackage{authblk}

\newcommand{\knorm}{{k_0 }} 
\newcommand{\amf}{{\omega }}

\newcommand{\p}[1]{{\left({#1}\right)}}
\newcommand{\comm}[1]{{\left[{#1}\right]}}
\newcommand{\acomm}[1]{{\left\{{#1}\right\}}}

\newcommand{\nn}{\nonumber}

\newcommand{\ads}{{\mathrm{AdS}_2}}
\newcommand{\mtwo}{{\mathscr{M}_2}}
\newcommand{\btwo}{{\mathscr{B}_2}}

\newcommand{\dd}{{\rm d}}
\newcommand{\ee}{{\rm e}}
\newcommand{\xx}{{\rm x}}
\newcommand{\im}{{\rm i}}

\numberwithin{equation}{section}

\title{Twisted Index on Hyperbolic Four-Manifolds}
\author{Daniele Iannotti and Antonio  Pittelli$^1$}
\affil{$\,^1$Dipartimento di Matematica, Universit\`a di Torino, \\ Via Carlo Alberto 10, 10123 Torino, Italy\\ INFN, Sezione di Torino, Via Pietro Giuria 1, 10125 Torino, Italy}

\begin{document}

\maketitle

\begin{abstract}
\noindent  
We introduce the topologically twisted index for four-dimensional $\mathcal N=1$ gauge theories quantized on $\ads \times S^1$.  We compute the index by applying supersymmetric localization to partition functions of vector and chiral multiplets  on $\ads \times T^2$, with and without a boundary: in both instances  we classify normalizability and boundary conditions  for gauge, matter and ghost fields. The index is twisted as the dynamical fields  are coupled to a  R-symmetry background 1-form with non-trivial exterior derivative and proportional to the spin connection.  After regularization the index is written in terms of  elliptic gamma functions, reminiscent of four-dimensional holomorphic blocks, and crucially  depends on the R-charge.

\end{abstract}

\newpage

\tableofcontents

\section{Introduction and Summary}

Supersymmetric field theories in both flat and curved spaces have been extensively studied over the years, serving as a crucial arena for advancing our theoretical understanding of quantum field theory (QFT), especially in the regime of strong interactions  \cite{Seiberg:1994rs,Intriligator:1996ex,Alday:2009aq}. While the complete information of a QFT  is contained in its generating functional of correlation functions, exact computations of this functional in interacting theories remain challenging. Nevertheless, the technique of supersymmetric localization \cite{Pestun:2007rz} has proven to be an extremely ductile  tool, enabling exact non-perturbative computations of specific generating functionals and other observables in a large class of  supersymmetric field theories defined on curved manifolds. In particular, localization techniques have been employed to study supersymmetric field theories on compact Riemannian manifolds, where a class of BPS observables can be precisely evaluated by reducing functional integrals to Gaussian integrals around a supersymmetric locus. Several such computations have been performed in various   dimension and for diverse topologies, leading to valuable insights  \cite{Hama:2011ea,Kapustin:2011jm,Kallen:2011ny,Hama:2012bg,Kallen:2012cs,Benini:2012ui,Alday:2013lba,Benini:2013xpa,Assel:2014paa,Benini:2015noa,Closset:2015rna,Panerai:2020boq,Festuccia:2020yff}, see also \cite{Pestun:2016zxk} and references therein.

Building on this success, this paper shifts attention towards studying supersymmetric gauge theories on non-compact hyperbolic manifolds,  focussing on $\ads \times T^2$, where by ${\rm AdS}_d$ we indicate $d$-dimensional  Anti-de Sitter space with Euclidean signature.  Gauge theories  in AdS have been investigated in connection with monodromy defects \cite{Giombi:2021uae}, black-hole entropy \cite{Cabo-Bizet:2017jsl,Rodriguez-Gomez:2017kxf},   chiral algebras \cite{Bonetti:2016nma} and   holomorphic blocks \cite{Beem:2012mb,Pittelli:2018rpl}. Moreover, the isometry group of ${\rm AdS}_d$ being the (global) conformal group in $\p{d-1}$-dimensions, QFT in AdS can be studied  via  conformal bootstrap    \cite{Carmi:2018qzm,Alday:2021ajh}.  Applying supersymmetric localization to QFTs on non-compact manifolds is also interesting from a technical viewpoint as  it requires the study of the behaviour at infinity of the degrees of freedom contributing to the path integral. This is necessary in order to make sure that not only  the final result for the partition function is convergent and well-defined, but also    supersymmetry is preserved. Alternatively, one can  consider a boundary at a specific distance from the origin of AdS and explore the interplay between supersymmetry,  boundary conditions and boundary degrees of freedom, as e.g. in \cite{Yoshida:2014ssa,Assel:2016pgi,David:2018pex,Longhi:2019hdh,Basu:2019mjo,David:2019ocd,David:2023btq,GonzalezLezcano:2023cuh,GonzalezLezcano:2023uar}.


In this  paper we present a detailed calculation of   partition functions for  $\mathcal N=1$ supersymmetric  gauge theories defined on $\ads \times T^2$. The construction of supersymmetric theories in a fixed background geometry involves taking a suitable limit of new minimal supergravity, leading to background fields coupled to a supersymmetric gauge theory with an R-symmetry, incorporating ordinary vector and chiral multiplets \cite{Festuccia:2011ws,Klare:2012gn,Dumitrescu:2012ha}. We  turn on a non-trivial R-symmetry background field  equal to  half the spin connection, which  is the usual setup corresponding to the topological twist, preserving on $\ads\times T^2$ two  Killing spinors of opposite chirality and R-charge. The metric we use on the four-manifold is refined by parameters that in the partition function combine into two complex moduli: one is a fugacity $p=e^{2 \pi \im \tau}$ for momentum on $T^2$, with $\tau$ being the torus modular parameter, while   the other is  a fugacity $q$ probing angular momentum on $\ads$. Such fugacities allow for linking partition functions  of supersymmetric  gauge theories  on $\ads \times T^2$   to  flavoured Witten indices for theories quantized on $\ads \times S^1$, namely 
\begin{align}\label{eq: ttionads2t2}
	 I_{\ads \times T^2} = {\rm Tr}_{\mathscr H}\comm{ \p{-1}^{\mathcal F} e^{- 2\pi \mathcal H}  } = {\rm Tr}_{\mathscr H}\comm{ \p{-1}^{\mathcal F} e^{- 2\pi \im \varphi_i {\mathcal Q}_i} q^{- {\mathcal  J} } p^{{\mathcal  P} }     } ~ ,
\end{align}
where ${\mathscr H}$ is the Hilbert space of BPS  states  on $\ads \times S^1$ whereas  the  operator $\mathcal H$ is the Hamiltonian, $\mathcal F$ is the fermionic number,  $\mathcal J$ is the angular momentum   on $\ads$,    $\mathcal P$ is the translation operator along   $S^1$    while     $\varphi_i$ and $ \mathcal Q_i  $ are the  chemical potential and the charge operator for the $i$-th flavour symmetry\footnote{Although the R-symmetry does not commute with the supercharges, the corresponding fugacity $\varphi_R$  appears in the index as  all other  flavour fugacities $\varphi_i$. Gauge fugacities appear as flavour fugacities that are integrated over.}, respectively.  Technically, supersymmetric  localization provides the plethystic exponential of  the single-letter  index (\ref{eq: ttionads2t2}). For instance,  let $\mathcal T$ be the gauge theory represented by  a dynamical vector multiplet in the adjoint representation of the gauge group $G$ coupled to a chiral multiplet of R-charge $r<1$ in a representation $\mathfrak R_G$ of $G$. The topologically twisted index of $\mathcal T$, defined on the four-dimensional hyperbolic manifold $\ads\times T^2$ in absence of boundaries, is given by the following integral over the Cartan torus of $G$ involving a ratio of elliptic Gamma functions:  
\begin{align}\label{eq: ttpfonads2t2example}
	  Z_{\ads \times T^2}  = \int \dd u   \prod_{\alpha \in {\rm Adj}\p{G}   }   \prod_{  \rho \in \mathfrak R_G } e^{2 \pi \im \widehat  \Psi  }  \frac{\Gamma_e\p{ e^{2 \pi \im (2 \varphi_R) }  e^{2 \pi \im \alpha\p{u} } ; q , p }   }{ \Gamma_e\p{ e^{2 \pi \im  \p{2-r} \varphi_R     } e^{ - 2 \pi \im \rho\p{u}   } ; q , p } } ~ , 
\end{align}
where  $\varphi_R$ and $u$ respectively are gauge and  R-symmetry fugacities. We refrain from explicitly writing down the  phase factor $\widehat \Psi$ as its form is not particularly illuminating; anyhow, $\widehat \Psi$ can be   read off from a suitable combination of the $\Psi^{\rm CM}_{D,R}$ and $\Psi^{\rm VM}_{D,R}$ reported in the main text.  We specified that (\ref{eq: ttpfonads2t2example}) is the expression valid for a chiral multiplet of R-charge lesser than one because the index involving the same multiplets  in the case $r\geq1$ has a   different form: the reason is that on non-compact $\ads\times T^2$ the normalizability of the fields contributing to the partition function dramatically depends on the R-charge. This phenomenon was already observed in lower dimension in relation to  the topologically twisted index computed via localization on $\ads \times S^1$ in absence of boundaries \cite{Pittelli:2018rpl}.

In presence of boundaries two dual sets of boundary conditions does not break supersymmetry: either Dirichlet conditions, requiring the vanishing of fields at the boundary;  or  Robin conditions,  requiring the vanishing of    derivatives of fields at the boundary. In fact,  derivatives are linear combinations of partial derivatives in directions that can be parallel and normal to the boundary, hence Robin conditions effectively are generalized Neumann conditions. A  boundary then allows for constructing  many different theories by just  combining multiplets satisfying a priori different  sets of supersymmetric boundary conditions. For example, (\ref{eq: ttpfonads2t2example}) can be interpreted as the twisted index of a gauge theory where Robin boundary conditions were imposed on the vector multiplet and Dirichlet boundary conditions were imposed upon the chiral multiplet. An intriguing  feature peculiar to the presence of boundaries is   how boundary degrees of freedom induce a \emph{flip} of   boundary conditions \cite{Dimofte:2017tpi,Longhi:2019hdh}: for instance,  suitably  coupling  a lower-dimensional matter multiplet to a four-dimensional chiral multiplet fulfilling  Dirichlet boundary conditions effectively yields a chiral multiplet with Robin boundary conditions. At the level of the partition function such a flip of boundary conditions is realized thanks to the multiplication properties of the special function appearing in the 1-loop determinants whose integral defines  the index:
 \begin{align}
   & Z_{\text{1-L}}^{\rm CM}|_R  = Z_{\text{1-L}}^{\rm CM}|_D  Z_{\text{1-L}}^{\rm CM}|_\partial      ~ , & Z_{\text{1-L}}^{\rm CM}|_\partial  = e^{2 \pi \im \Psi^{\rm CM}_\partial  } /\theta_0\p{ e^{2 \pi \im q_i \gamma_i } ; q } ~ , 
\end{align}
where $\theta_0\p{z,q} = \p{z,q}_\infty\p{q/z ; q}_\infty$, with $\p{z;q}_\infty = \prod_{j\geq0}\p{ 1 - z q^j}$ being the indefinite $q$-Pochhammer symbol,   $Z_{\text{1-L}}^{\rm CM}|_{D,R}$ are the four-dimensional chiral-multiplet 1-loop determinants of fluctuations  satisfying  Dirichlet and Robin boundary conditions, respectively, while $Z_{\text{1-L}}^{\rm CM}|_{\partial}$ is the 	1-loop determinant of the three-dimensional boundary multiplet.

  

In summary, generalizing localization techniques to the case of  non-compact manifolds   naturally  opens up new avenues for exploration. Compelling future directions include a careful  study of the phase factors $\Psi^{\rm CM}_{D, R}$ and $\Psi^{\rm VM}_{D, R}$ appearing after regularization of the 1-loop determinants for  chiral and vector multiplets, as such phases encode important scheme-independent informations about   anomalies, vacuum energy and central charges of the corresponding gauge theory in hyperbolic spacetime \cite{Benini:2012cz,Cassani:2013dba,DiPietro:2014bca,Assel:2014paa,Assel:2015nca}.   Furthermore, it would be very interesting to analyze the large-$N$ limit of $\mathcal N=4$ supersymmetric Yang-Mills theory with gauge group $SU(N)$ on $\ads \times T^2$ as such limit should unveil BPS configurations similar to the black-strings  in ${\rm AdS}_5$  detected in the topologically twisted index on the compact manifold $S^2 \times T^2$ \cite{Closset:2013sxa,Hosseini:2016cyf,Hosseini:2019lkt,Hosseini:2020vgl}. In particular, we predict that the  constraint
 \begin{align}
   & \gamma_R  - \frac{\amf}{2}  + \alpha_x \frac{\tau}{2}     =  \frac{\alpha_y }{2}  ~ , & \alpha_x , \alpha_y \in \mathbb Z ~ , 
\end{align}
 which we derive in the main text, should also appear in the   dual gravity theory with $\gamma_R$, $\omega$  being related to the  electrostatic potential and the angular velocity   of the supergravity solution, respectively. Besides, it would be very intriguing to investigate  non-perturbative dualities  for gauge theories on $\ads\times T^2$; especially in relation  of boundary degrees of freedom, which are known to be affected by such transformations in a non-trivial way \cite{Gaiotto:2008ak}. Eventually, possible generalizations of this paper comprehend the addiction of BPS defects,  vortices or orbifold structures \cite{Alday:2012au,Imamura:2012rq,Imamura:2013qxa,Inglese:2023wky} on $\ads \times T^2$, as all these objects yield   further refinements of  the index \cite{Drukker:2012sr}.


\paragraph{Outline.}

In Section \ref{sec: susybackground} we set up the background geometry by introducing the chosen metric and frame on $\ads \times T^2$. We then find its  rigid supersymmetric completion by solving   the conformal   Killing spinor equation  on $\ads \times T^2$ endowed with a   background field for the $U(1)_R$ R-symmetry that is  proportional to the spin connection and encodes the topological twist. Thus, we show that such conformal Killing spinors  also solve the Killing spinor equation with a suitable choice of background fields descending from new minimal supergravity. Hence, we study periodicities and global smoothness of Killing spinors on topologically twisted $\ads \times T^2$. In Section \ref{sec: susycohomology} we write down the supersymmetric multiplets involved in our analysis, their supersymmetry  variations in component fields and Lagrangians. Then, we rewrite the supersymmetry transformations in cohomological form by introducing a new set of fields that makes manifest the fundamental degrees of freedom contributing to the partition function. Moreover, we show how supersymmetric boundary conditions emerge from either supersymmetry-exact deformations of the Lagrangian or the  equations of motion. In Section \ref{sec: susylocalization} we calculate the path integral of topologically twisted gauge theories on  $\ads \times T^2$ by means of supersymmetric localization. We first solve the BPS equations for  vector and chiral multiplets,  thus obtaining the supersymmetric locus over which dynamical fields fluctuate. Then, we  calculate the contribution to the partition function of such fluctuations, giving rise to a non-trivial 1-loop determinant expressed as an infinite product that can be regularized  in terms of special functions. We explicitly display the various possibilities corresponding to different choices of either  boundary or normalizability  conditions imposed on supersymmetric fluctuations.




 \section{Supersymmetric Background}\label{sec: susybackground}


We   choose the following line element  on $\ads  \times T^2$:
\begin{align}\label{eq: lineelementreal}
    \dd s^2 = L^2  \dd \theta^2 + L^2  \sinh^2\theta\p{ \dd \varphi + \Omega_3 \dd x + \Omega_4 \dd y }^2 + L^2 \beta^2\comm{ \p{\dd x + \tau_1 \dd y }^2 + \tau_2^2 \dd y^2  } ~ ,
\end{align} 
where the four-dimensional metric $g_{\mu\nu}$ can be read off from the usual relation $\dd s^2 = g_{\mu \nu} \dd \xx^\mu \dd \xx^\nu $, with $\xx^\mu = \p{ \theta , \varphi , x , y} $. In particular, $\theta\in [0, +\infty)$ and $\varphi \in [0, 2 \pi )$ are  coordinates on $\ads $, while  $x , y \in [0, 2 \pi )$ are coordinates on $T^2$. The parameter $L$ has dimension of length and encodes the radius of $\ads$ appearing e.g. in the Ricci scalar $R_\ads = - 2/L^2$.  The dimensionless  parameters $\Omega_1, \Omega_2 \in \mathbb R$ introduce in the partition function of the theory a fugacity for the angular momentum on $\ads$, as in \cite{Closset:2013sxa,Benini:2015noa}; whereas $\tau_1, \tau_2 \in \mathbb R$ respectively are real and imaginary  part of the modular parameter $\tau= \tau_1 + \im \tau_2$ of the torus  $T^2$. Finally, the dimensionless parameter   $\beta \in \mathbb R$ parametrizes the scale of $T^2$ with respect to the radius of   $\ads$. We shall also  consider a boundary at $\theta = \theta_0 >0$ to  explore the interplay between bounday conditions and boundary degrees of freedom.

We adopt the orthonormal frame
\begin{align}\label{eq: frameortho}
    & \ee^1 = L \dd \theta  ~ , &  \ee^2 = L \sinh \theta\p{ \dd\varphi + \Omega_3 \dd x + \Omega_4 \dd y }  ~ , \nn\\
    & \ee^3 = L \beta \p{ \dd x + \tau_1 \dd  y }  ~ ,  & \ee^4 = L \beta \tau_2 \dd y ~ ,
\end{align}
satisfying e.g.  $g_{\mu \nu} = \delta_{a b} \ee^a_\mu \ee^b_\nu$  and  $\delta^{a b} = g^{\mu \nu} \ee^a_\mu \ee^b_\nu$. In the frame (\ref{eq: frameortho}) the  non-trivial components of the spin connection read
\begin{align}
    \omega_{12}  = - \omega_{21} = - \cosh \theta \p{ \dd \varphi + \Omega_3 \dd x + \Omega_4 \dd y  } ~ .
\end{align}
On $\ads \times T^2$ the conformal Killing-spinor equations, 
\begin{align}\label{eq:CKSE}
     \p{ \nabla_\mu - \im A^C_\mu  } \zeta + \frac{1}{4} \sigma_\mu \widetilde\sigma^\nu \p{ \nabla_\nu - \im A^C_\nu  } \zeta = 0 ~ , \nn\\
     \p{ \nabla_\mu + \im A^C_\mu  } \widetilde \zeta + \frac{1}{4} \widetilde  \sigma_\mu \sigma^\nu \p{ \nabla_\nu + \im A^C_\nu  } \widetilde  \zeta = 0 ~ ,
\end{align}
is solved by  
\begin{align}\label{eq: ks}
    & \zeta_\alpha = \sqrt{\knorm } \, e^{ \frac{\im}{2}\p{ \alpha_\varphi \varphi + \alpha_x x  + \alpha_y y   } } \begin{pmatrix} 1 \\ 0 \end{pmatrix}_\alpha  ~ , & \widetilde \zeta^{\dot \alpha } =  \sqrt{\knorm } \, e^{ -  \frac{\im}{2}\p{ \alpha_\varphi \varphi + \alpha_x x  + \alpha_y y   } } \begin{pmatrix} 0 \\ 1 \end{pmatrix}^{\dot \alpha }  ~ ,
\end{align}
where  $k_0 \in \mathbb C$ is  a normalization constant and $\alpha_{2,3,4} \in \mathbb R$ parametrize  non-trivial phases of $\zeta, \widetilde\zeta$ along the three circles inside $\ads \times T^2$, while $A^C$ is the   background field
\begin{align}
    A^C = \frac{1}{2} \p{  \omega_{12} + \alpha_\varphi \dd\varphi + \alpha_x \dd x + \alpha_y \dd y } ~ .
\end{align}
Moreover, the spinors (\ref{eq: ks}) fulfil the Killing-spinor equations
\begin{align}\label{eq: kse}
    \p{ \nabla_\mu - \im A_\mu  }\zeta + \im V_\mu \zeta + \im V^\nu \sigma_{\mu \nu} \zeta = 0  ~ , \nn\\
    \p{ \nabla_\mu + \im A_\mu  } \widetilde \zeta - \im V_\mu \widetilde  \zeta - \im V^\nu \widetilde  \sigma_{\mu \nu}  \widetilde  \zeta = 0  ~ , 
\end{align}
with background fields
\begin{align}
    V & = L \beta \kappa \p{ \dd x + \tau \dd y  }   ~ , \nn\\
    A & = A^C + \frac{3}{2} V ~ ,
\end{align}
where $\kappa$ is an arbitrary constant  and  the 1-forms $A^C$ and $A^R$ are smooth on  $\ads$ if $\alpha_\varphi = 1$. Thus,  the $\zeta$ and $\widetilde \zeta $ reported in (\ref{eq: ks}) are Killing spinors of R-charge $\pm 1$, respectively. As the field strength $F^{(R)}$ of the R-symmetry   field is non-trivial and satisfies   $F^{(R)} = \dd A = \p{1/2} \dd \omega_{12}$, the Killing spinors $\zeta$ and $\widetilde \zeta $ describe a supersymmetric $\ads \times T^2$ background with a topological twist on $\ads$, analogous to those investigated  in the case of compact manifolds e.g. in  four  \cite{Closset:2013sxa}  and  three    \cite{Benini:2015noa} dimensions. On a compact two-dimensional manifold $\mtwo$   the direct link between $F^{(R)}$ and $\dd \omega_{12}$ characterizing the topological twist implies that the R-symmetry flux equals  the Euler characteristic of $\mtwo$, up to a sign. On a two-dimensional manifold with boundary $\btwo$ the R-symmetry flux $\mathfrak f_R$   is proportional to the line integral of $A$ along the one-dimensional boundary $\partial \btwo $. Indeed,  applying Stokes' theorem to the smooth   R-symmetry field   $A_{(0)} = A|_{\alpha_\varphi = 1}$    gives 
\begin{align}
   \mathfrak f_R = \frac{1}{2 \pi } \int_\btwo    F^{(R)} = \frac{1}{2 \pi } \int_\btwo   \dd A_{(0)} = \frac{1}{2 \pi } \oint_{S^1_0 } A_{(0)} = \frac{1}{2}\p{1- \cosh \theta_0}    ~ ,
\end{align}
with $\btwo$ being $\ads$ with a boundary  at $\theta=\theta_0$ and $S^1_0=\partial \btwo$ being the circle in $\ads$ at $\theta=\theta_0$. As observed in \cite{Cabo-Bizet:2017jsl}, on $\ads$   fluxes   are not quantized, as opposed to what happens e.g. on the two-sphere, where the single-valuedness of   transition functions between different patches requires all fluxes to take integer values.

Imposing either periodicity or anti-periodicity of the Killing spinors $\zeta, \widetilde \zeta$ along the torus circles parametrized by $x$ and $y$ yields 
\begin{align}
    \alpha_x, \alpha_y \in \mathbb Z ~ .
\end{align}
Furthermore, smoothness of the R-symmetry field $A$ requires $\alpha_\varphi=1$, implying the anti-periodicity of the Killing spinors along the shrinking circle in $\ads$ parametrized by   $\varphi$. The Killing spinors reported in (\ref{eq: ks}) are  manifestly smooth in every point of the four-manifold apart from the origin as $\zeta, \widetilde \zeta$ are written in the frame (\ref{eq: frameortho}), which is singular  at $\theta = 0$ due to $ \varphi$ being undefined at the origin. Smoothness at $\theta=0$ can be examined  by first  rotating  (\ref{eq: frameortho}) into a frame that is non-singular at the origin via a local Lorentz transformation ${\ell^a}_b$,
\begin{align}
    & \delta {\ee^a}_\mu = - {\ell^a}_b {\ee^b}_\mu  ~ , & {\ell^a}_b = { \begin{pmatrix}
        0 & \varphi \\
        - \varphi & 0 
    \end{pmatrix}^a}_b ~ ,
\end{align}
which in turn induces the following rotation upon $\zeta , \widetilde \zeta$:
\begin{align}
    \zeta' & = \mathfrak L^{-1} \zeta  = \sqrt{\knorm } \, e^{ \frac{\im}{2}\comm{ \p{\alpha_\varphi - 1 } \varphi + \alpha_x x  + \alpha_y y   } } \begin{pmatrix} 1 \\ 0 \end{pmatrix}  ~ , \nn\\
    \widetilde \zeta' & = \widetilde{\mathfrak L}^{-1} \widetilde \zeta  =  \sqrt{\knorm } \, e^{ -  \frac{\im}{2}\comm{ \p{\alpha_\varphi - 1 } \varphi + \alpha_x x  + \alpha_y y   } } \begin{pmatrix} 0 \\ 1 \end{pmatrix}   ~ ,
\end{align}
where  
\begin{align}
    & \mathfrak L = \exp\p{ - \frac{1}{2}\ell_{ ab }\sigma^{ab}} ~ , & \widetilde {\mathfrak L} = \exp\p{ - \frac{1}{2}\ell_{ ab }\widetilde\sigma^{ab}} ~ ,
\end{align}
encode the action of local Lorentz transformations upon left- and right-handed spinors, respectively. The spinors $\zeta', \widetilde \zeta'$ are independent of the coordinate $\varphi$ if and only if $\alpha_\varphi=1$, which is then the value making  the Killing spinors smooth on the whole four-manifold.


\section{Supersymmetry and Cohomology}\label{sec: susycohomology}

\subsection{Vector Multiplet} 

 
A  vector multiplet   enjoying  $\mathcal N=1$ supersymmetry consists  of  a 1-form $a_\mu $ encoding the gauge field, two   complex
spinors $\lambda , \widetilde \lambda $ of opposite chirality parametrizing the gauginos and a  0-form $D$ corresponding to an auxiliary field ensuring off-shell closure of the supersymmetry algebra. The fields $\p{a_\mu , \lambda, \widetilde \lambda , D}$ have R-charges $\p{0, +1 , -1 , 0}$,  transform  in the adjoint
representation of the gauge group $G$. The  vector-multiplet  supersymmetry variations with respect to  $\zeta, \widetilde \zeta $ read 
\begin{align}\label{eq: vm_susy}
\delta a_\mu & = \im \widetilde \zeta \widetilde  \sigma_\mu \lambda   + \im \zeta \sigma_\mu \widetilde \lambda   ~ , \nn \\
\delta \lambda  & = f_{\mu \nu} \sigma^{\mu\nu} \zeta + \im D \zeta   ~ , \nn \\
\delta \widetilde \lambda  & = f_{\mu \nu} \widetilde \sigma^{\mu\nu} \widetilde \zeta - \im D \widetilde \zeta    ~ , \nn \\
\delta D & = \widetilde \zeta \widetilde\sigma^\mu \p{ D_\mu \lambda + \frac{3 \im }{2} V_\mu \lambda } -  \zeta \sigma^\mu \p{ D_\mu \widetilde \lambda - \frac{3 \im }{2} V_\mu \widetilde \lambda } ~ ,
\end{align}
where $f$ is the field strength of the gauge field $a$ with components $f_{\mu \nu} = \partial_\mu a_\nu - \partial_\nu a_\mu - \im q_G \comm{a_\mu , a_\nu} $. The constant $q_G$ is the gauge charge appearing in the covariant derivative
\begin{align}
    D_\mu = \nabla_\mu - \im q_R A_\mu  - \im q_G a_\mu \circ_{\mathfrak R_G}   ~ ,
\end{align}
where $\circ_{\mathfrak R_G}$ represents  the action upon a field $\Phi$ in the representation $\mathfrak R_G$ of the gauge group $G$. The bosonic fields $a_\mu$ and $D$ of the vector multiplet satisfy the reality conditions
\begin{align}\label{eq: vmrc}
    & a_\mu^\dagger  = a_\mu   ~ ,  & D^\dagger  =  - D   ~ ,
\end{align}
whereas  there is no need to impose reality conditions upon the vector-multiplet fermionic fields $\lambda, \widetilde \lambda$. The supersymmetry transformations (\ref{eq: vm_susy}) can be rewritten in cohomological form as follows:
\begin{align}\label{eq: vmcc}
   & \delta a = \Lambda  ~ , & \delta \Lambda  = 2 \im \p{ \mathcal L_K + \mathcal G_{\Phi_G} } a  ~ ,    \nn\\
   & \delta \Phi_G =  0  ~ ,    \nn\\
   & \delta \Psi    = \Delta   ~ , & \delta \Delta   = 2 \im \p{ \mathcal L_K + \mathcal G_{\Phi_G} } \Psi   ~ , 
\end{align}
where we introduced the Grassmann-even 0-forms $\Phi_G, \Delta $ as well as the Grassmann-odd 0-form $\Psi$ and 1-form $\Lambda_\mu$ given by  
\begin{align}
    & \Phi_G = \iota_K a  ~ , &  \Delta = D - 2 \im Y^\mu \widetilde Y^\nu f_{\mu\nu } ~ , \nn\\
    & \Lambda_\mu =  \im \widetilde \zeta \widetilde  \sigma_\mu \lambda   + \im \zeta \sigma_\mu \widetilde \lambda    ~ , & \Psi   =  \frac{\zeta^\dagger \lambda}{2 \im |\zeta|^2 } - \frac{\widetilde \zeta^\dagger \widetilde  \lambda}{2 \im |\widetilde \zeta|^2 }   ~ , \nn\\
    & \lambda_\alpha   = \im\p{ \Psi - \iota_{\widetilde K} \Lambda  } \zeta_\alpha + \im \frac{\iota_Y \Lambda }{|\zeta|^2} \zeta^\dagger_\alpha       ~ , &  \widetilde\lambda^{\dot\alpha } = -  \im\p{ \Psi + \iota_{\widetilde K} \Lambda  } \widetilde\zeta^{\dot\alpha} - \im \frac{\iota_{\widetilde Y} \Lambda }{|\widetilde \zeta|^2} \widetilde \zeta^{\dagger\dot \alpha}     ~ , 
\end{align}
with
\begin{align}\label{eq:KV}
    K^\mu = \zeta \sigma^\mu \widetilde \zeta ~ , \qquad Y^\mu = \frac{   \zeta \sigma^\mu \widetilde \zeta^\dagger  }{2  |\widetilde \zeta|^2 } ~ , \qquad \widetilde Y^\mu = - \frac{ \zeta^\dagger \sigma^\mu \widetilde \zeta  }{2 |\zeta|^2   } ~ ,  \qquad \widetilde K^\mu = \frac{ \widetilde  \zeta^\dagger \sigma^\mu \widetilde \zeta^\dagger   }{4 |\zeta|^2  |\widetilde \zeta|^2   } ~ , 
\end{align}
being the Killing-spinor bilinears defined in \cite{Closset:2013sxa}. The norms $|\zeta|^2$ and $|\widetilde \zeta|^2$ descend  from the  complex conjugates of the Killing spinors, which are 
\begin{align} 
    & \zeta^{\dagger \alpha} = \p{ \zeta_\alpha }^* = \sqrt{ \knorm^* } \, e^{ - \frac{\im}{2}\p{ \alpha_\varphi \varphi + \alpha_x x  + \alpha_y y   } } \begin{pmatrix} 1 \\ 0 \end{pmatrix}^\alpha  ~ , & |\zeta|^2 = |\knorm|  ~ , \nn\\ 
    & \widetilde \zeta^\dagger_{\dot \alpha } = \p{ \widetilde \zeta^{\dot \alpha } }^* =  \sqrt{\knorm^* } \, e^{   \frac{\im}{2}\p{ \alpha_\varphi \varphi + \alpha_x x  + \alpha_y y   } } \begin{pmatrix} 0 \\ 1 \end{pmatrix}_{\dot \alpha }  ~ ,  & |\widetilde\zeta|^2 = |\knorm|  ~ ,
\end{align}
providing in turn the reality conditions on  Killing-spinor bilinears:
\begin{align} 
    & K_\mu^*  = 4 |\zeta|^2 |\widetilde \zeta|^2 \widetilde K^\mu   ~ ,  & Y_\mu^* = \frac{ | \zeta|^2 }{ |\widetilde \zeta|^2} \widetilde Y^\mu   ~ . 
\end{align}
In particular, the Killing-spinor equations (\ref{eq: kse}) imply that  $K^\mu$, which in our setup reads 
\begin{align}
    K^\mu \partial_\mu  = \frac{\im \knorm}{L \beta \tau_2}\p{  \amf \partial_\varphi  - \tau \partial_x + \partial_y  } ~ ,
\end{align} 
is a Killing vector:
\begin{align}
    \nabla_\mu K_\nu = \im  \epsilon_{\mu \nu  \lambda  \rho  }  K^\lambda V^\rho \qquad \to \qquad  \nabla_{(\mu }K_{\nu)} = 0 ~ .
\end{align}
In  (\ref{eq: vmcc})  the supersymmetry variation $\delta$ manifests itself as an equivariant differential  fulfilling  
\begin{align}
   \delta^2 = 2 \im \p{ \mathcal L_K + \mathcal G_{\Phi_G} } ~ ,
\end{align}
where the Lie derivative $\mathcal L_K$ generates  a spacetime isometry of the manifold while 
\begin{align}\label{eq: phigaction}
     \mathcal G_{\Phi_G} a & = - \dd_a \Phi_G = -  \dd \Phi_G + \im q_G \comm{a , \Phi_G }   ~ , \nn\\
     \mathcal G_{\Phi_G} X & = -  \im q_G   \Phi_G \circ_{\mathfrak R_G} X    ~ ,  \qquad X \neq a ~ , 
\end{align}
represent the action of gauge transformations upon fields. In the case of weakly gauged theories with background vector multiplets,  (\ref{eq: phigaction}) is interpreted as the action of the flavour group $G\equiv G_F$.



The vector-multiplet Lagrangian,
\begin{align}
      \mathcal L_{\rm VM} = \frac{1}{4} f_{\mu \nu} f^{\mu \nu } - \frac{1}{2}D^2 + \frac{\im}{2}\lambda \sigma^\mu D_\mu \widetilde \lambda + \frac{\im}{2} \widetilde \lambda \widetilde \sigma^\mu D_\mu   \lambda - \frac{3}{2} V_\mu \widetilde \lambda \widetilde \sigma^\mu \lambda   ~ , 
\end{align}
is $\delta$-exact up to boundary terms,
\begin{align}\label{eq: vmlagexact}
      \mathcal L_{\rm VM} & = \delta  V_{\rm VM} + \frac{\im }{2}\nabla_\mu\comm{ \widetilde Y^\mu \p{  \lambda  \lambda }  - Y^\mu \p{ \widetilde \lambda \widetilde \lambda }  }  ~ , \nn\\
      & = \delta  V_{\rm VM} + \im  \nabla_\mu\comm{  \widetilde Y^\mu \p{ \Psi - \iota_{\widetilde K} \Lambda  }\iota_Y \Lambda   -  Y^\mu \p{ \Psi + \iota_{\widetilde K} \Lambda  }\iota_{\widetilde Y} \Lambda   }  ~ , 
\end{align}
with deformation term
\begin{align}
      V_{\rm VM} & =  \frac{1}{4|\zeta|^2} \p{\delta \lambda_\alpha}^\dagger \lambda_\alpha + \frac{1}{4|\widetilde\zeta|^2} \p{\delta \widetilde \lambda^{\dot\alpha} }^\dagger \widetilde  \lambda^{\dot\alpha}   ~ ,  \nn\\
      & = \frac{1}{2}\p{\delta \Psi }^\dagger \Psi + \frac{1}{2}\p{\iota_{\widetilde K} \delta \Lambda  }^\dagger \iota_{ \widetilde K}  \Lambda  + \frac{1}{4 |\zeta|^4}\p{\iota_{Y} \delta \Lambda  }^\dagger \iota_{Y}  \Lambda + \frac{1}{4 |\widetilde\zeta|^4}\p{\iota_{\widetilde Y} \delta \Lambda  }^\dagger \iota_{\widetilde Y}  \Lambda    ~ .
\end{align}
In absence of boundaries the total derivative in (\ref{eq: vmlagexact}) is irrelevant and the corresponding action $\mathcal S_{\rm CM}$ is $\delta$-exact and then manifestly supersymmetric. In presence of boundaries the total-derivative terms in (\ref{eq: vmlagexact}) drop out if the following dual sets of supersymmetric boundary conditions are imposed:  
\begin{align}
   & {\rm Robin} : & \iota_Y a|_\partial = \iota_{\widetilde Y} a|_\partial    = \iota_Y \Lambda|_\partial = \iota_{\widetilde Y} \Lambda|_\partial = 0 ~ , \nn\\
   & {\rm Dirichlet} : & \iota_K a|_\partial = \iota_{\widetilde K} a|_\partial    = \Psi_\partial   = \iota_{\widetilde K} \Lambda|_\partial = 0 ~ ,  
\end{align}
together with the vanishing of the corresponding supersymmetry variations. Especially, Dirichlet conditions only affect the components of the gauge field $a_\mu$ that are parallel to the boundary, whereas Robin conditions mix with each other components that are either parallel or orthogonal to the boundary. After including   Faddeev-Popov ghosts $c, \widetilde c$ and their supersymmetric completion, as  e.g. in  \cite{Pestun:2007rz,Assel:2016pgi},  the BRST-improved supersymmetry variation $\p{\delta + \delta_{\rm BRST}}a_\mu = \p{ \Lambda_\mu +  D_\mu c }$  implies 
\begin{align}
   & {\rm Robin} : &   L_{\widetilde Y} c|_\partial =  L_{ Y} c|_\partial =   L_{\widetilde Y} \widetilde c|_\partial =  L_{ Y} \widetilde c|_\partial = 0 ~ , \nn\\
   & {\rm Dirichlet} : & c|_\partial = \widetilde c|_\partial =  0 ~ .
\end{align}


\subsection{Chiral multiplet}


$\mathcal N=1$  chiral multiplets in a representation $\mathfrak R_G$ of the gauge group $G$  contain a  0-form $\phi  $, a   left-handed  spinor $\psi $ and  a  0-form $F$, where the latter is a non-dynamical  field that, in analogy with $D$, allows the closure of the supersymmetry algebra on the chiral multiplet  without using the equations of motion. The fields $\p{ \phi  , \psi , F }$, whose  R-charges are $\p{r , r - 1  , r - 2 }$, are related to each other by the following  supersymmetry variations: 
\begin{align}\label{eq: cm_susy}
    \delta \phi & =  \sqrt{2} \zeta \psi  ~ , \nn\\
    \delta \psi & = \sqrt{2} F \zeta + \im \sqrt{2} \p{\sigma^\mu \widetilde \zeta }  D_\mu \phi ~ , \nn\\
    \delta F & =  \im \sqrt{2} \widetilde\zeta \widetilde\sigma^\mu \p{ D_\mu \psi - \frac{\im}{2} V_\mu \psi } - 2 \im \p{\widetilde\zeta \widetilde \lambda} \phi   ~ . 
\end{align}
 The bosonic fields $\phi $ and $F$ of the chiral multiplet fulfil  the reality conditions
\begin{align}
    &  \phi^\dagger   = \phi    ~ ,  & F^\dagger  =  - \widetilde F  ~ ,
\end{align}
where $\widetilde \phi$ and $\widetilde F$, together with the right-handed spinor $\widetilde \psi$, form an anti-chiral multiplet\footnote{In Euclidean spacetime imposing reality conditions upon     fermionic fields is not necessary.} in the conjugate representation $\overline{ \mathfrak R}_G$ of the gauge group $G$. Their supersymmetry transformation reads 
\begin{align}\label{eq: acm_susy}
    \delta \widetilde \phi & =  \sqrt{2}  \widetilde  \zeta  \widetilde  \psi  ~ , \nn\\
    \delta  \widetilde  \psi & = \sqrt{2}  \widetilde  F  \widetilde  \zeta + \im \sqrt{2} \p{ \widetilde  \sigma^\mu  \zeta }  D_\mu   \widetilde  \phi ~ , \nn\\
    \delta  \widetilde  F & =  \im \sqrt{2}   \widetilde  \zeta \sigma^\mu \p{ D_\mu  \widetilde  \psi + \frac{\im}{2} V_\mu   \widetilde  \psi } + 2 \im \widetilde  \phi  \p{ \zeta  \lambda}    ~ .
\end{align}
If we define cohomological fields correspoding to the Grassmann-odd 0-forms $B,C$ and the Grassmann-even 0-form $\Xi$
\begin{align}
    & B =  \frac{\zeta^\dagger  \psi}{\sqrt 2 |\zeta|^2}  ~ , & C = \sqrt{2} \zeta \psi  ~ , \nn\\
    & \psi  =  \sqrt{2} B \zeta - \frac{C \zeta^\dagger }{\sqrt{2} |\zeta|^2 } ~ , & \Xi   =  F - 2 \im L_{\widetilde Y} \phi   ~ ,  
\end{align}
 where
\begin{align}
   & L_v =  v^\mu D_\mu   ~ , & v \in \acomm{ K, \widetilde K, Y , \widetilde Y}  ~ ,
\end{align}
is the covariant Lie derivative along the vector $v$, the relations (\ref{eq: cm_susy}) can be written in cohomological form:
 \begin{align}\label{eq: cmcc}
     &  \delta X_i = X'_i ~ , &  \delta X_i' = 2 \im \p{  \mathcal L_K - \im q_R \Phi_R + \mathcal G_{\Phi_G}  } X_i  ~ , 
 \end{align}
 with $i=1,2$, where  $X_1 = \phi$,  $X_2 = B$, $X_1' = C $ and $X_2' = \Xi$. The supersymmetry variation of the auxiliary field $ F$ in cohomological form  is 
 \begin{align}
    \delta F = 2 \im \p{ L_K B + L_{\widetilde Y} C  - \im \iota_{\widetilde Y}\Lambda  \phi  } ~ .
 \end{align}
 The structure of (\ref{eq: cmcc}) implies that the supersymmetry variation $\delta $ behaves as an equivariant differential also on chiral-multiplet fields, where $\Phi_R$ is the R-symmetry counterpart of the 0-form $\Phi_G$,
\begin{align}
    \Phi_R = \iota_K A = \frac{\im \knorm}{2 L \beta \tau_2}\p{   \amf \alpha_\varphi  - \tau \alpha_x + \alpha_y  }  ~ .
\end{align}
More generally,  $\delta^2$ acts  upon a field $X$ of R-charge $q_R$, flavour charge $q_F$ and gauge charge $q_G$ as
 \begin{align}\label{eq: delta2flavoured}
     \delta^2  X = 2 \im L_K X = 2 \im \p{  \mathcal L_K  + \mathcal G_{\Phi_R}    + \mathcal G_{\Phi_F}  + \mathcal G_{\Phi_G}  } X  ~ ,
 \end{align}
where 
 \begin{align}
    & \mathcal G_{\Phi_R} X = - \im q_R \Phi_R X  ~ ,  &\Phi_R = \iota_K A = \frac{\im \knorm}{2 L \beta \tau_2}\p{   \amf \alpha_\varphi  - \tau \alpha_x + \alpha_y  }  ~ ,
\end{align}
while $\mathcal G_{\Phi_F}$ formally acts  as in (\ref{eq: phigaction}) with respect to the flavour group $G_F$. The object $\Phi_R$ did not appear in  (\ref{eq: vmcc}) as the fields $\p{a ,  \Lambda , \Phi_G , \Psi , \Delta  }$ are R-symmetry neutral. The supersymmetry variations (\ref{eq: acm_susy}) can be recast in the form reported in (\ref{eq: cmcc}) by defining
\begin{align}
    & \widetilde B =  \frac{\widetilde \zeta^\dagger  \widetilde \psi}{\sqrt 2 |\widetilde \zeta|^2}  ~ , & \widetilde  C = \sqrt{2} \widetilde  \zeta \widetilde  \psi  ~ , \nn\\
    & \widetilde  \psi  =  \sqrt{2} \widetilde  B \widetilde  \zeta - \frac{\widetilde  C \widetilde  \zeta^\dagger }{\sqrt{2} |\widetilde \zeta|^2 } ~ , & \widetilde \Xi   = \widetilde   F + 2 \im L_{ Y} \widetilde  \phi   ~ ,
\end{align}
and choosing $X_1 =  \widetilde\phi$,  $X_2 = \widetilde B$, $X_1' = \widetilde C $ as well as $X_2' = \widetilde \Xi$. In cohomological form the supersymmetry variation of the auxiliary field $\widetilde F$  reads
\begin{align}
    \delta \widetilde F = 2 \im \p{ L_K \widetilde B - L_{ Y} \widetilde C  - \im \widetilde\phi  \iota_{ Y}  \Lambda    } ~ .
 \end{align}



The chiral-multiplet Lagrangian, 
\begin{align}
    \mathcal L_{\rm CM} & = D_\mu \widetilde \phi D^\mu \phi  + \im V^\mu \comm{ \p{D_\mu \widetilde \phi} \phi - \widetilde \phi D_\mu \phi } + \p{r/4} \widetilde \phi \p{R + 6 V^2 + D} \phi - \widetilde F F \nn\\
    & + \im \widetilde \psi \widetilde \sigma^\mu D_\mu \psi + \p{V^\mu/2} \widetilde \psi \widetilde \sigma_\mu \psi + \im \sqrt{2}\p{ \widetilde \phi \lambda \psi - \widetilde \psi \widetilde \lambda \phi } ~ , 
\end{align}
which in cohomological fields reads 
\begin{align}
    \mathcal L_{\rm CM} & = 4  L_K \widetilde \phi L_{\widetilde K} \phi + 4 L_Y \widetilde \phi L_{\widetilde Y}\phi   -  \p{ \widetilde \Xi - 2 \im L_Y \widetilde \phi  } \p{ \Xi + 2 \im L_{\widetilde Y} \phi  } \nn\\
    & + 2  \im \p{ \kappa \widetilde \phi L_K \phi  + \widetilde B L_K B + \widetilde B L_{\widetilde Y} C - \widetilde C L_Y B + \widetilde C L_{\widetilde K} C }  - \kappa \widetilde C C \nn\\
    &   + q_G  \comm{  \widetilde C \iota_{\widetilde K} \Lambda  \phi - 2 \widetilde \phi \iota_Y \Lambda B + 2 \widetilde B \iota_{\widetilde Y} \Lambda \phi +  \widetilde C \Psi \phi + \widetilde \phi \p{\Delta  - 2 \im  \iota_{\widetilde K}  L_K a    } \phi -  \widetilde \phi \p{\Psi - \iota_{\widetilde K} \Lambda   } C  }   ~ ,
\end{align}
is $\delta$-exact with respect to the deformation term $V_{\rm CM}$ given by
\begin{align}
     & \mathcal L_{\rm CM} = \delta  V_{\rm CM}  ~ ,  & V_{\rm CM} = - 2 \im \widetilde B L_{\widetilde Y} \phi - 2 \im \widetilde C L_{\widetilde K} \phi - \widetilde F B + q_G \widetilde \phi \p{\Psi  - \iota_{\widetilde K}\Lambda  }\phi - \kappa \widetilde \phi C  ~ .
\end{align}
Consequently,
\begin{align}
    \delta \mathcal L_{\rm CM} = 2 \im \mathcal L_K V_{\rm CM} = 2 \im \nabla_\mu \p{ K^\mu  V_{\rm CM} } ~ . 
\end{align}
In absence of boundaries  the supersymmetry variation of $\mathcal L_{\rm CM}$ being a total derivative readily implies that the variation of the corresponding action $\mathcal S_{\rm CM}$ is zero. Moreover,  since $K^\theta=0$, in presence of a boundary at $\theta=\theta_0$  the Lagrangian of the chiral multiplet is supersymmetric for any choice of  boundary conditions,  where the latter can be  obtained  by imposing the vanishing of the boundary terms generated by the equations of motion for to the bulk Lagrangian: 
\begin{align}
    \delta_{\rm eom} \mathcal L_{\rm CM} & = \p{ {\rm bulk} } +  4 \nabla_\mu \comm{  Y^\mu  \delta_{\rm eom}  \widetilde \phi L_{\widetilde Y} \phi + 4  \widetilde Y^\mu  \p{ L_Y  \widetilde \phi }  \delta_{\rm eom} \phi } \nn\\
    & + 2 \im  \nabla_\mu  \p{  \widetilde Y^\mu \widetilde B  \delta_{\rm eom}  C - Y^\mu \widetilde C \delta_{\rm eom}  B  }    ~ , 
\end{align}
 where the bulk terms vanish on the solution of the equations of motion. The boundary terms descending from the equations of motion cancel out if the following dual sets  of supersymmetric boundary conditions are imposed:
\begin{align}
    & {\rm Dirichlet} :  & \phi_\partial =\widetilde \phi_\partial = C_\partial =\widetilde C_\partial = 0 ~ , \nn\\
    & {\rm Robin} :  & B_\partial =\widetilde B_\partial = \p{L_{\widetilde Y}\phi}_\partial = \p{L_{ Y}\widetilde\phi}_\partial = 0 ~ . 
\end{align}


\section{Supersymmetric Localization}\label{sec: susylocalization}

We now compute the partition function of gauge theories coupled to matter via supersymmetric localization \cite{Pestun:2007rz}. We focus on Abelian gauge theories as the generalization to the non-Abelian case is straightforward. We start by deriving the supersymmetric locus solving the BPS equations; then, we will compute the 1-loop determinant of the fluctuations over the BPS locus.

\subsection{BPS Locus} 

 The vector-multiplet BPS equations  are
\begin{align}
    \lambda = \delta \lambda = \widetilde \lambda = \delta \widetilde \lambda = 0 ~ , 
\end{align}
which in cohomological form read 
\begin{align}\label{eq: vmbpsequations}
     & \p{ \mathcal L_K + \mathcal G_{\Phi_G} } a = 0 ~ ,  &  D - 2 \im Y^\mu \widetilde Y^\nu f_{\mu \nu } = 0 ~ . 
\end{align}
We employ the following ansatz: 
\begin{align}\label{eq: vmagansatz}
    a = \comm{ a_\varphi\p{\theta} + b_\varphi } \dd \varphi +  \comm{ a_x\p{\theta} + b_x }  \dd x + \comm{ a_y\p{\theta} + b_y }  \dd y   ~ , 
\end{align}
where we set to zero the pure-gauge component $a_x\p{\theta}$, while  $a_\varphi, a_x, a_y  $ are complex functions and   the flat connections $b_\varphi, b_x, b_y $ are complex constants, a priori. The gauge field above is smooth on $\ads$ if $b_\varphi = - a_\varphi\p{0}$. By plugging the ansatz (\ref{eq: vmagansatz}) into the BPS equations (\ref{eq: vmbpsequations}) we obtain the complex BPS locus for the vector multiplet:
\begin{align}\label{eq: vmbpslocuscomplex}
    \comm{ \amf a_\varphi\p{\theta} - \tau a_x\p{\theta}  + a_y\p{\theta } }_{\rm BPS } = a_0 = \text{constant} ~ ,  \nn\\
    D|_{\rm BPS } = L^{-2} \sinh^{-1}\p{\theta} a'_\varphi\p{\theta}   ~ ,
\end{align}
 implying that the BPS value of the gauge fugacity $\Phi_G$ is manifestly constant,
\begin{align} 
     \Phi_G|_{\rm BPS} = \frac{ \im \knorm  }{L \beta \tau_2} \p{ a_0 +  \amf  b_\varphi - \tau b_x + b_y }  ~ .
\end{align}
Imposing the reality  conditions reported in (\ref{eq: vmrc}) yields the real BPS locus     
\begin{align}\label{eq: vmbpslocusreal}
        & a_\varphi\p{\theta}|_{\rm BPS} = a_\varphi = {\rm constant} ~ , \nn\\
        & a_x\p{\theta}|_{\rm BPS} = a_x = {\rm constant} ~ , \nn\\
        & a_y\p{\theta}|_{\rm BPS} = a_y = {\rm constant} ~ , \nn\\
        & D|_{\rm BPS} = 0 ~ , \nn\\
        & {\rm Re} \p{ \amf } a_\varphi  - \tau_1 a_x  + a_y  = {\rm Re}  \p{a_0}  ~ , \nn\\
        & {\rm Im} \p{ \amf } a_\varphi  - \tau_2 a_x    = {\rm Im}  \p{a_0}  ~ .
\end{align}
Using (\ref{eq: vmbpslocusreal}) for a smooth connection on $\ads$ gives $b_\varphi = - a_\varphi $   and    
\begin{align} 
     \Phi_G|_{\rm BPS} = \frac{ \im \knorm  }{L \beta \tau_2} \comm{     a_y   + b_y - \tau \p{ a_x  + b_x}  }  ~ .
\end{align}
In presence of a boundary at $\theta = \theta_0$ there are two possibilities: if Dirichlet conditions are imposed, then $\Phi_G|_{\rm BPS}$ has to vanish at the boundary. Since $\Phi_G|_{\rm BPS}$ is constant, Dirichlet conditions require  $\Phi_G|_{\rm BPS}=0$ everywhere. Instead, Robin boundary conditions do not impose any constraint on $\Phi_G = \iota_K a$, which hence stays non trivial.

The BPS equations for the chiral multiplet read
\begin{align}
    \psi  = \delta \psi  = \widetilde \psi  = \delta \widetilde \psi  = 0 ~ , 
\end{align}
which in cohomological form are 
\begin{align}\label{eq: cmbpsequations}
     &\p{  \mathcal L_K - \im q_R \Phi_R - \im q_G \Phi_G   } \phi = 0  ~ , \nn\\
     & \p{  \mathcal L_K + \im q_R \Phi_R + \im q_G \Phi_G   } \widetilde \phi = 0  ~ , \nn\\
    &  F = 2 \im L_{\widetilde Y} \phi    ~ , \nn\\
     & \widetilde  F = - \, 2 \im L_{ Y}  \widetilde \phi  ~ ,
 \end{align}
 with $\phi=\phi\p{\theta, \varphi, x , y}$ and $\widetilde \phi=\widetilde \phi\p{\theta, \varphi, x , y}$ being periodic in $\varphi, x, y$. For generic values of $\Phi_R, \Phi_G$  the trivial locus 
 \begin{align}\label{eq: cmbpslocus}
 \phi = \widetilde \phi = F = \widetilde F =  0 ~ ,
 \end{align}
 is the only solution to  (\ref{eq: cmbpsequations}, regardless of the presence of a boundary at $\theta=\theta_0$. In particular, the value of fields reported in (\ref{eq: cmbpslocus}) trivializes   the classical contribution to the partition function given e.g. by superpotential terms.


\subsection{One-Loop Determinant}

The 1-loop determinant of supersymmetric fluctuations over the BPS locus for a chiral multiplet is
\begin{align}\label{eq: cmoneloopdeterminantmasterformula}
   Z_{\text{1-L}}^{\rm CM} = \frac{\det_{{\rm Ker} L_{Y}} \delta^2}{\det_{{\rm Ker} L_{\widetilde Y}} \delta^2} ~ ,
\end{align}
with the kernel of the differential operator $L_Y = Y^\mu D_\mu$ being spanned by functions $B_{m_\varphi , m_x , m_y} $ labelled by integers corresponding to the Fourier modes around the  circles parametrized by $\varphi, x, y$:
\begin{align}
     & {\rm Ker} L_{Y} : B_{m_\varphi , m_x , m_y}  = e^{\im m_\varphi \varphi + \im  m_x x +  \im  m_y y } B^{(0)}_{m_\varphi , m_x , m_y}\p{\theta}  ~ , &  m_\varphi , m_x , m_y \in \mathbb Z ~ .
\end{align}
The behaviour of the modes $B_{m_\varphi , m_x , m_y}$ at the origin of $\ads$  is
\begin{align}
    \lim_{\theta \to 0 } B^{(0)}_{m_\varphi , m_x , m_y}\p{\theta} \sim  \theta^{m_\varphi } ~ ,
\end{align}
meaning that $B_{m_\varphi , m_x , m_y}$ is non-singular at $\theta = 0$ if  $m_\varphi   \in \mathbb N$. The modes $B_{m_\varphi , m_x , m_y} $ satisfy  the eigenvalue equation $\delta^2 B_{m_\varphi , m_x , m_y}  = \lambda_B B_{m_\varphi , m_x , m_y} ~ , $ where 
\begin{align}
     \lambda_B & =  \frac{ 2 \im \knorm}{L \beta \tau_2}\comm{ - \amf m_\varphi + \tau m_x - m_y + q_G \p{a_0 + b_y - \tau b_x - \amf a_\varphi } + \frac{r-2}{2}\p{ \amf - \tau \alpha_x + \alpha_y } }  ~ ,
\end{align}
is the eigenvalue contributing to the numerator of $Z_{\text{1-L}}^{\rm CM}$. Analogously,   the kernel of the differential operator $L_{\widetilde Y} = {\widetilde Y}^\mu D_\mu$ is  spanned by functions $\phi_{n_\varphi , n_x , n_y} $,
\begin{align}
     & {\rm Ker} L_{\widetilde Y} : \phi_{n_\varphi , n_x , n_y}  = e^{\im n_\varphi \varphi + \im  n_x x +  \im  n_y y } \phi^{(0)}_{n_\varphi , n_x , n_y}\p{\theta } ~ , &  n_\varphi , n_x , n_y \in \mathbb Z ~ ,
\end{align}
whose behaviour near the origin of $\ads$ is
\begin{align}
    \lim_{\theta \to 0 } \phi^{(0)}_{n_\varphi , n_x , n_y}\p{\theta} \sim  \theta^{- n_\varphi } ~ ,
\end{align}
implying that $\phi_{n_\varphi , n_x , n_y} $ is non singular at $\theta=0$ if $\p{-n_\varphi } = \ell_\varphi \in \mathbb N$. The modes $\phi_{n_\varphi , n_x , n_y} $ are eigenfunctions of the operator $\delta^2$ with eigenvalue 
\begin{align}
     \lambda_\phi & =  \frac{ 2 \im \knorm}{L \beta \tau_2}\comm{  \amf \ell_\varphi + \tau n_x - n_y + q_G \p{a_0 + b_y - \tau b_x - \amf a_\varphi } + \frac{r}{2}\p{ \amf - \tau \alpha_x + \alpha_y } }  ~ ,
\end{align}
which contributes  to the denominator of $Z_{\text{1-L}}^{\rm CM} $. In both $\lambda_B$ and $\lambda_\phi$ the R-charges $q_R^\phi = r$, $q_R^B = \p{r-2}$ as well as the gauge charge $q_G$ respectively multiply the same quantities $\gamma_R$ and $ \gamma_G$, where  
\begin{align}\label{eq: fugacities}
   \gamma_R & = \frac{1}{2}\p{ \amf - \tau \alpha_x + \alpha_y }  ~ , \nn\\
   \gamma_G & = a_0  - \amf a_\varphi - \tau b_x + b_y    ~  .
\end{align}
Especially, the first line in (\ref{eq: fugacities}) can  be interpreted  as a constraint on the chemical potentials $\gamma_R, \omega, \tau$, as  in the case of gauge theories on $S^3 \times S^1$ dual to ${\rm AdS}_5$ black holes \cite{Cabo-Bizet:2018ehj}.

In presence of boundaries we have two possible 1-loop determinants: on the one hand, if we impose Dirichlet conditions, the modes $\phi_{n_\varphi , n_x , n_y}$ have to satisfy a first-order homogeneous  differential equation with boundary condition $\phi_{n_\varphi , n_x , n_y}|_\partial=0$,  implying $\phi_{n_\varphi , n_x , n_y}=0$. Therefore, Dirichlet conditions kill the modes $\phi_{n_\varphi , n_x , n_y}$ contributing to the denominator of $Z_{\text{1-L}}^{\rm CM} $, leaving the modes $B_{m_\varphi , m_x , m_y}$ unaffected. The result is
\begin{align}
   Z_{\text{1-L}}^{\rm CM}|_D = \prod_{m_\varphi \in \mathbb N} \prod_{m_x , m_y \in \mathbb Z} \comm{   \amf m_\varphi - \tau m_x + m_y - q_G \gamma_G - \p{r-2} \gamma_R }  ~ ,
\end{align}
which can be regularized by means of (\ref{eq: reginfprodtriple}), yielding
\begin{align}
   Z_{\text{1-L}}^{\rm CM}|_D & = e^{ 2 \pi \im \Psi^{\rm CM}_{D} } / \Gamma_e\p{ e^{2 \pi \im \p{ 2 \gamma_R - q_i \gamma_i}  } ; q , p }    ~ , \nn\\
   \Psi^{\rm CM}_{D} & =   \frac{1}{24  \tau  \omega  }\p{1 +  \tau  - 2  \p{ 2 \gamma_R - q_i \gamma_i} + \omega  } \nn\\
   & \times \comm{ 2  \p{ 2 \gamma_R - q_i \gamma_i} \p{ \p{ 2 \gamma_R - q_i \gamma_i} - \omega  - 1} + \omega  +  \tau \p{1 - 2 \p{ 2 \gamma_R - q_i \gamma_i} + \omega  } }   ~ ,
\end{align}
with $q_i \gamma_i = \p{ r \gamma_R + q_G \gamma_G}$. On the other hand, if we impose Robin conditions, the modes $B_{m_\varphi , m_x , m_y}$ have to satisfy a first-order homogeneous  differential equation with boundary condition $B_{m_\varphi , m_x , m_y}|_\partial=0$,  which sets   $ B_{m_\varphi , m_x , m_y} =0$ everywhere. As a consequence, Robin conditions trivialize the modes $B_{m_\varphi , m_x , m_y}$ contributing to the numerator  of the chiral-multiplet 1-loop determinant and leave the modes  $\phi_{n_\varphi , n_x , n_y}$   untouched because $L_{\widetilde Y}\phi_{n_\varphi , n_x , n_y}=0$ on the whole four-manifold by definition. Thus,
\begin{align}
   Z_{\text{1-L}}^{\rm CM}|_R = \prod_{\ell_\varphi \in \mathbb N} \prod_{n_x , n_y \in \mathbb Z} \comm{   \amf \ell_\varphi + \tau n_x - n_y + q_G \gamma_G + r \gamma_R }^{-1} ~ ,
\end{align}
whose regularized form provided by (\ref{eq: reginfprodtriple}) reads
\begin{align}
   Z_{\text{1-L}}^{\rm CM}|_R & = e^{ 2 \pi \im \Psi^{\rm CM}_{R} } \Gamma_e\p{ e^{2 \pi \im  q_i \gamma_i  } ; q , p }    ~ , \nn\\
   \Psi^{\rm CM}_{R} & =  - \frac{1}{24  \tau  \omega  }\p{1 +  \tau  - 2 q_i \gamma_i + \omega  }\comm{ 2 q_i \gamma_i \p{ q_i \gamma_i - \omega  - 1} + \omega  +  \tau \p{1 - 2 q_i \gamma_i + \omega  } }   ~ . 
\end{align}
As observed e.g. in \cite{Dimofte:2017tpi,Longhi:2019hdh}, dual 1-loop determinants are mapped to each other by multiplication of 1-loop determinants corresponding ot boundary multiplets: 
\begin{align}
   & Z_{\text{1-L}}^{\rm CM}|_R  = Z_{\text{1-L}}^{\rm CM}|_D  Z_{\text{1-L}}^{\rm CM}|_\partial      ~ , & Z_{\text{1-L}}^{\rm CM}|_\partial  = e^{2 \pi \im \Psi^{\rm CM}_\partial  } /\theta_0\p{ e^{2 \pi \im q_i \gamma_i } ; q } ~ , \nn\\
   & Z_{\text{1-L}}^{\rm CM}|_D  = Z_{\text{1-L}}^{\rm CM}|_R  Z_{\text{1-L}}^{\rm FM}|_\partial      ~ , & Z_{\text{1-L}}^{\rm FM}|_\partial  = e^{-2 \pi \im \Psi^{\rm CM}_\partial  }  \theta_0\p{ e^{2 \pi \im q_i \gamma_i } ; q } ~ , 
\end{align}
with the boundary phase being given by
\begin{align}
    \Psi^{\rm CM}_\partial = - \frac{1}{12 \omega } \comm{1 + 6 q_i \gamma_i \p{ q_j \gamma_j  - 1 - \omega  } + \omega  \p{3 + \omega } } = \Psi^{\rm CM}_R - \Psi^{\rm CM}_D  ~ .
\end{align}

Eventually, in absence of boundaries, we require that both $\phi_{n_\varphi , n_x , n_y}$ and $B_{m_\varphi , m_x , m_y}$ are square integrable on $\ads \times T^2$ according to the integral measure 
\begin{align}
   \int_{\ads \times T^2} \sqrt{\det g} \, |\Phi |^2  ~ .
\end{align}
By inspection, the behaviour of the modes $B_{m_\varphi , m_x , m_y}$ at infinity is
\begin{align}
    \lim_{\theta \to +\infty  } B^{(0)}_{m_\varphi , m_x , m_y}\p{\theta} \sim  e^{\theta \p{r - 2}/2 } ~ ,
\end{align}
and taking into account that 
\begin{align}
    \lim_{\theta \to +\infty  }  \sqrt{\det g} \sim  e^{\theta  } ~ ,
\end{align}
as well as that the modes  are finite near the origin at $\theta\sim0$, normalizability of $B_{m_\varphi , m_x , m_y}$ requires
\begin{align}
   r < 1 ~ .
\end{align}
Analogously, the behaviour of the modes $\phi_{n_\varphi , n_x , n_y}$ at infinity is 
\begin{align}
    \lim_{\theta \to + \infty } \phi^{(0)}_{n_\varphi , n_x , n_y}\p{\theta} \sim  e^{ - \theta \,  r /2} ~ ,
\end{align}
and normalizability of  $\phi_{n_\varphi , n_x , n_y}$ imposes 
\begin{align}
   r > 1 ~ .
\end{align}
In summary, 
\begin{align}\label{eq: cmoneloopdeterminantnoboundarynormalizable}
   & Z_{\text{1-L}}^{\rm CM}|_{r<1}  = Z_{\text{1-L}}^{\rm CM}|_D     ~ , & Z_{\text{1-L}}^{\rm CM}|_{r>1}  = Z_{\text{1-L}}^{\rm CM}|_R     ~ ,
\end{align}
and $Z_{\text{1-L}}^{\rm CM}|_{r=1}=1$ as there are no normalizable modes for $r=1$.

Similarly to what happens in three dimensions \cite{Assel:2016pgi}, the 1-loop determinant for a non-Abelian  vector multiplet enjoying $\mathcal N=1$ supersymmetry is 
\begin{align}\label{eq: vmoneloopdeterminantmasterformula}
     Z_{\text{1-L}}^{\text{VM} }  = \sqrt{\frac{ \p{\det_\Psi  \delta^2 }\p{ \det_c \delta^2 } \p{ \det_{\widetilde c} \delta^2 } }{ \p{ \det_{\iota_{\widetilde K} a}  \delta^2 } \p{ \det_{\iota_Y a} \delta^2 } \p{ \det_{\iota_{\widetilde Y} a} \delta^2 }  } } ~ ,
\end{align}
where $\Psi, c$ and $\widetilde c$ contribute as modes $B_{m_\varphi , m_x , m_y}$, in the adjoint representation of the gauge group $G$, with R-charge $r=2$, while $ \iota_{\widetilde Y} a , \iota_{\widetilde K} a$ and $\iota_{ Y} a$ contribute as modes $\phi_{n_\varphi , n_x , n_y}$ in the adjoint of $G$ with R-charges $\p{2, 0 , -2}$, respectively. Nonetheless, if Dirichlet conditions upon vector multiplet modes are imposed, only  $ \iota_{\widetilde Y} a , \iota_{ Y} a$ survive and after simplifications we find
\begin{align}
     Z_{\text{1-L}}^{\text{VM} }|_{\rm D}  =  \frac{ 1 }{  \det_{\iota_Y \mathcal A}  \delta^2    }  =   \p{  {\det}_{\phi }  \delta^2    }^{-1}_{r=2} = \p{ Z_{\text{1-L}}^{\rm CM}|_R }_{r=2} ~ ,
\end{align}
where the product over roots of the adjoint representation of $G$ is understood. On the other hand,  Robin  conditions kill $ \iota_{\widetilde Y} a , \iota_{ Y} a$, leaving the other modes invariant; therefore, after a few other  simplifications,
\begin{align}
     Z_{\text{1-L}}^{\text{VM} }|_{\rm R}  =    {\det}_c  \delta^2 =  \p{  {\det}_B  \delta^2 }_{r=2} = \p{ Z_{\text{1-L}}^{\rm CM}|_D }_{r=2}     ~ .
\end{align}

Instead, in absence of boundaries, all modes appearing in (\ref{eq: vmoneloopdeterminantmasterformula}) do contribute, a priori. In fact, various contributions  drop  out, giving at the end of the day
\begin{align} 
     Z_{\text{1-L}}^{\text{VM} }  = \frac{  \det_c \delta^2   }{  \det_{\iota_Y a} \delta^2    } = \p{Z_{\text{1-L}}^{\text{CM} }}_{r=2} = \p{ Z_{\text{1-L}}^{\rm CM}|_R }_{r=2}   ~ ,
\end{align}
where  the last equality holds if we restrict to normalizable modes only, as in (\ref{eq: cmoneloopdeterminantnoboundarynormalizable}).


\appendix

\section{Regularization of Infinite Products}

In the main text we found that  1-loop determinants of supersymmetric multiplets on $\ads \times T^2$ are triple infinite products of the form
\begin{align}
           Q( b_0 | a_0 , c_0 )_\infty  &  = \prod_{\ell \in \mathbb N}  \prod_{n_1 , n_2 \in \mathbb Z}  \p{   a_0 n_1 +  n_2 + c_0 \ell  + b_0 }   ~ ,
\end{align}
with $a_0, b_0$ and $c_0$ being complex constants. The product above can be regularized by   rewriting the products over $\mathbb Z$ as double products over $\mathbb N$ and   using multiple Zeta and Gamma functions \cite{FRIEDMAN2004362}:
\begin{align}
           Q_{s_1 , s_2}( b | a_{12} , c )_\infty &  = \prod_{\ell \in \mathbb N} \prod_{n_1 , n_2 \in \mathbb N} \p{   a_{12} n_1 +  n_2 + c \ell +  b } \p{  - a_{12} n_1 +  n_2 + 1  -   c \ell -  b } \nn\\
          & \times  \comm{   - a_{12} \p{ n_1 + 1} +  n_2 + c \ell +  b } \comm{   a_{12} \p{ n_1 + 1} +  n_2 + 1  -   c \ell -  b }   ~ , \nn\\    
          & =  \frac{1}{\Gamma_3\p{  b | c , a_{12} , 1  } \Gamma_3\p{ 1 - b | - c , - a_{12} , 1  } \Gamma_3\p{  b - a_{12} | c, - a_{12} , 1  } \Gamma_3\p{  1 - b + a_{12} | - c , a_{12} , 1  } }  ~ ,     \nn\\
          & = e^{\im \pi \comm{ \zeta_3\p{ 0 , b | c , a_{12} , 1  } + \zeta_3\p{ 0 , 1 - b + a_{12} | - c , a_{12} , 1  } } } \nn\\
          & \times \p{ e^{2 \pi \im b  } ; e^{2 \pi \im c} , e^{2 \pi \im a_{12}} }_\infty  \p{ e^{2 \pi \im \p{ a_{12} - b}  } ; e^{ - 2 \pi \im c} , e^{2 \pi \im a_{12}} }_\infty   ~ , \nn\\     
          & = e^{\im \pi \comm{ \zeta_3\p{ 0 , b | c , a_{12} , 1  } + \zeta_3\p{ 0 , 1 - b + a_{12} | - c , a_{12} , 1  } } } \frac{ \p{ e^{2 \pi \im b  } ; e^{2 \pi \im c} , e^{2 \pi \im a_{12}} }_\infty }{ \p{ e^{2 \pi \im \p{ a_{12} + c - b}  } ; e^{  2 \pi \im c} , e^{2 \pi \im a_{12}} }_\infty }  ~ , \nn\\     
          & = e^{\im \pi \comm{ \zeta_3\p{ 0 , b | c , a_{12} , 1  } + \zeta_3\p{ 0 , 1 - b + a_{12} | - c , a_{12} , 1  } } } /   \Gamma_e\p{ e^{2 \pi \im b  } ; e^{2 \pi \im c} , e^{2 \pi \im a_{12}} }   ~ ,    
\end{align}
where   
\begin{align}
   \Gamma_e\p{z , q_1 , q_2 }  = \frac{\p{ q_1 q_2 /z ; q_1 ; q_2}_\infty }{ \p{  z ; q_1 ; q_2}_\infty }  ~ ,
\end{align}
is the elliptic Gamma function. Altogether,
\begin{align}\label{eq: reginfprodtriple}
           \prod_{\ell \in \mathbb N}  \prod_{n_1 , n_2 \in \mathbb Z}  \frac{1}{   a_0 n_1 +  n_2 + c_0 \ell  + b_0 } \to   e^{ 2 \pi \im \Psi_{(3)}\p{a_{12} , b , c } } \Gamma_e\p{ e^{2 \pi \im b  } ; e^{2 \pi \im c} , e^{2 \pi \im a_{12}} }    ~ , \nn\\
          \Psi_{(3)}\p{a_{12} , b , c } =  - \frac{1}{24 a_{12} c }\p{1 + a_{12} - 2 b + c }\comm{ 2 b \p{ b - c - 1} + c + a_{12}\p{1 - 2 b + c } }   ~ . 
\end{align}


\bibliographystyle{unsrt}
\bibliography{4d_N1_susy}

\begin{thebibliography}{10}

\bibitem{Seiberg:1994rs}
N.~Seiberg and Edward Witten.
\newblock {Electric - magnetic duality, monopole condensation, and confinement
  in N=2 supersymmetric Yang-Mills theory}.
\newblock {\em Nucl. Phys. B}, 426:19--52, 1994.
\newblock [Erratum: Nucl.Phys.B 430, 485--486 (1994)].

\bibitem{Intriligator:1996ex}
Kenneth~A. Intriligator and N.~Seiberg.
\newblock {Mirror symmetry in three-dimensional gauge theories}.
\newblock {\em Phys. Lett. B}, 387:513--519, 1996.

\bibitem{Alday:2009aq}
Luis~F. Alday, Davide Gaiotto, and Yuji Tachikawa.
\newblock {Liouville Correlation Functions from Four-dimensional Gauge
  Theories}.
\newblock {\em Lett. Math. Phys.}, 91:167--197, 2010.

\bibitem{Pestun:2007rz}
Vasily Pestun.
\newblock {Localization of gauge theory on a four-sphere and supersymmetric
  Wilson loops}.
\newblock {\em Commun. Math. Phys.}, 313:71--129, 2012.

\bibitem{Hama:2011ea}
Naofumi Hama, Kazuo Hosomichi, and Sungjay Lee.
\newblock {SUSY Gauge Theories on Squashed Three-Spheres}.
\newblock {\em JHEP}, 05:014, 2011.

\bibitem{Kapustin:2011jm}
Anton Kapustin and Brian Willett.
\newblock {Generalized Superconformal Index for Three Dimensional Field
  Theories}.
\newblock 6 2011.

\bibitem{Kallen:2011ny}
Johan Kallen.
\newblock {Cohomological localization of Chern-Simons theory}.
\newblock {\em JHEP}, 08:008, 2011.

\bibitem{Hama:2012bg}
Naofumi Hama and Kazuo Hosomichi.
\newblock {Seiberg-Witten Theories on Ellipsoids}.
\newblock {\em JHEP}, 09:033, 2012.
\newblock [Addendum: JHEP 10, 051 (2012)].

\bibitem{Kallen:2012cs}
Johan K\"all\'en and Maxim Zabzine.
\newblock {Twisted supersymmetric 5D Yang-Mills theory and contact geometry}.
\newblock {\em JHEP}, 05:125, 2012.

\bibitem{Benini:2012ui}
Francesco Benini and Stefano Cremonesi.
\newblock {Partition Functions of ${\mathcal{N}=(2,2)}$ Gauge Theories on
  S$^{2}$ and Vortices}.
\newblock {\em Commun. Math. Phys.}, 334(3):1483--1527, 2015.

\bibitem{Alday:2013lba}
Luis~F. Alday, Dario Martelli, Paul Richmond, and James Sparks.
\newblock {Localization on Three-Manifolds}.
\newblock {\em JHEP}, 10:095, 2013.

\bibitem{Benini:2013xpa}
Francesco Benini, Richard Eager, Kentaro Hori, and Yuji Tachikawa.
\newblock {Elliptic Genera of 2d ${\mathcal{N}}$ = 2 Gauge Theories}.
\newblock {\em Commun. Math. Phys.}, 333(3):1241--1286, 2015.

\bibitem{Assel:2014paa}
Benjamin Assel, Davide Cassani, and Dario Martelli.
\newblock {Localization on Hopf surfaces}.
\newblock {\em JHEP}, 08:123, 2014.

\bibitem{Benini:2015noa}
Francesco Benini and Alberto Zaffaroni.
\newblock {A topologically twisted index for three-dimensional supersymmetric
  theories}.
\newblock {\em JHEP}, 07:127, 2015.

\bibitem{Closset:2015rna}
Cyril Closset, Stefano Cremonesi, and Daniel~S. Park.
\newblock {The equivariant A-twist and gauged linear sigma models on the
  two-sphere}.
\newblock {\em JHEP}, 06:076, 2015.

\bibitem{Panerai:2020boq}
Rodolfo Panerai, Antonio Pittelli, and Konstantina Polydorou.
\newblock {Topological Correlators and Surface Defects from Equivariant
  Cohomology}.
\newblock {\em JHEP}, 09:185, 2020.

\bibitem{Festuccia:2020yff}
Guido Festuccia, Anastasios Gorantis, Antonio Pittelli, Konstantina Polydorou,
  and Lorenzo Ruggeri.
\newblock {Cohomological localization of $ \mathcal{N} $ = 2 gauge theories
  with matter}.
\newblock {\em JHEP}, 09:133, 2020.

\bibitem{Pestun:2016zxk}
Vasily Pestun et~al.
\newblock {Localization techniques in quantum field theories}.
\newblock {\em J. Phys. A}, 50(44):440301, 2017.

\bibitem{Giombi:2021uae}
Simone Giombi, Elizabeth Helfenberger, Ziming Ji, and Himanshu Khanchandani.
\newblock {Monodromy defects from hyperbolic space}.
\newblock {\em JHEP}, 02:041, 2022.

\bibitem{Cabo-Bizet:2017jsl}
Alejandro Cabo-Bizet, Victor~I. Giraldo-Rivera, and Leopoldo~A. Pando~Zayas.
\newblock {Microstate counting of AdS$_{4}$ hyperbolic black hole entropy via
  the topologically twisted index}.
\newblock {\em JHEP}, 08:023, 2017.

\bibitem{Rodriguez-Gomez:2017kxf}
Diego Rodriguez-Gomez and Jorge~G. Russo.
\newblock {Free energy and boundary anomalies on $\mathbb{S}^a\times
  \mathbb{H}^b$ spaces}.
\newblock {\em JHEP}, 10:084, 2017.

\bibitem{Bonetti:2016nma}
Federico Bonetti and Leonardo Rastelli.
\newblock {Supersymmetric localization in AdS$_{5}$ and the protected chiral
  algebra}.
\newblock {\em JHEP}, 08:098, 2018.

\bibitem{Beem:2012mb}
Christopher Beem, Tudor Dimofte, and Sara Pasquetti.
\newblock {Holomorphic Blocks in Three Dimensions}.
\newblock {\em JHEP}, 12:177, 2014.

\bibitem{Pittelli:2018rpl}
Antonio Pittelli.
\newblock {Supersymmetric localization of refined chiral multiplets on
  topologically twisted $H^2$ \texttimes{} $S^1$}.
\newblock {\em Phys. Lett. B}, 801:135154, 2020.

\bibitem{Carmi:2018qzm}
Dean Carmi, Lorenzo Di~Pietro, and Shota Komatsu.
\newblock {A Study of Quantum Field Theories in AdS at Finite Coupling}.
\newblock {\em JHEP}, 01:200, 2019.

\bibitem{Alday:2021ajh}
Luis~F. Alday, Agnese Bissi, and Xinan Zhou.
\newblock {One-loop gluon amplitudes in AdS}.
\newblock {\em JHEP}, 02:105, 2022.

\bibitem{Yoshida:2014ssa}
Yutaka Yoshida and Katsuyuki Sugiyama.
\newblock {Localization of three-dimensional $\mathcal{N}=2$ supersymmetric
  theories on $S^1 \times D^2$}.
\newblock {\em PTEP}, 2020(11):113B02, 2020.

\bibitem{David:2018pex}
Justin~R. David, Edi Gava, Rajesh~Kumar Gupta, and Kumar Narain.
\newblock {Boundary conditions and localization on AdS. Part I}.
\newblock {\em JHEP}, 09:063, 2018.

\bibitem{Longhi:2019hdh}
Pietro Longhi, Fabrizio Nieri, and Antonio Pittelli.
\newblock {Localization of 4d $\mathcal{N}=1$ theories on $\mathbb{D}^2\times
  \mathbb{T}^2$}.
\newblock {\em JHEP}, 12:147, 2019.

\bibitem{Basu:2019mjo}
Rudranil Basu and Augniva Ray.
\newblock {Supersymmetric Localization on dS: Sum over topologies}.
\newblock {\em Eur. Phys. J. C}, 80(9):885, 2020.

\bibitem{David:2019ocd}
Justin~R. David, Edi Gava, Rajesh~Kumar Gupta, and Kumar Narain.
\newblock {Boundary conditions and localization on AdS. Part II. General
  analysis}.
\newblock {\em JHEP}, 02:139, 2020.

\bibitem{David:2023btq}
Justin~R. David, Edi Gava, Rajesh~Kumar Gupta, and K.~S. Narain.
\newblock {Discontinuities of free theories on $AdS_2$}.
\newblock 3 2023.

\bibitem{GonzalezLezcano:2023cuh}
Alfredo Gonz\'alez~Lezcano, Imtak Jeon, and Augniva Ray.
\newblock {Supersymmetric localization: \ensuremath{\mathscr{N}} = (2) theories
  on S$^{2}$ and AdS$_{2}$}.
\newblock {\em JHEP}, 07:056, 2023.

\bibitem{GonzalezLezcano:2023uar}
Alfredo Gonz\'alez~Lezcano, Imtak Jeon, and Augniva Ray.
\newblock {Supersymmetry and complexified spectrum on Euclidean AdS$_2$}.
\newblock 5 2023.

\bibitem{Festuccia:2011ws}
Guido Festuccia and Nathan Seiberg.
\newblock {Rigid Supersymmetric Theories in Curved Superspace}.
\newblock {\em JHEP}, 06:114, 2011.

\bibitem{Klare:2012gn}
Claudius Klare, Alessandro Tomasiello, and Alberto Zaffaroni.
\newblock {Supersymmetry on Curved Spaces and Holography}.
\newblock {\em JHEP}, 08:061, 2012.

\bibitem{Dumitrescu:2012ha}
Thomas~T. Dumitrescu, Guido Festuccia, and Nathan Seiberg.
\newblock {Exploring Curved Superspace}.
\newblock {\em JHEP}, 08:141, 2012.

\bibitem{Dimofte:2017tpi}
Tudor Dimofte, Davide Gaiotto, and Natalie~M. Paquette.
\newblock {Dual boundary conditions in 3d SCFT\textquoteright{}s}.
\newblock {\em JHEP}, 05:060, 2018.

\bibitem{Benini:2012cz}
Francesco Benini and Nikolay Bobev.
\newblock {Exact two-dimensional superconformal R-symmetry and
  c-extremization}.
\newblock {\em Phys. Rev. Lett.}, 110(6):061601, 2013.

\bibitem{Cassani:2013dba}
Davide Cassani and Dario Martelli.
\newblock {Supersymmetry on curved spaces and superconformal anomalies}.
\newblock {\em JHEP}, 10:025, 2013.

\bibitem{DiPietro:2014bca}
Lorenzo Di~Pietro and Zohar Komargodski.
\newblock {Cardy formulae for SUSY theories in $d =$ 4 and $d =$ 6}.
\newblock {\em JHEP}, 12:031, 2014.

\bibitem{Assel:2015nca}
Benjamin Assel, Davide Cassani, Lorenzo Di~Pietro, Zohar Komargodski, Jakob
  Lorenzen, and Dario Martelli.
\newblock {The Casimir Energy in Curved Space and its Supersymmetric
  Counterpart}.
\newblock {\em JHEP}, 07:043, 2015.

\bibitem{Closset:2013sxa}
Cyril Closset and Itamar Shamir.
\newblock {The $\mathcal{N}=1$ Chiral Multiplet on $T^2\times S^2$ and
  Supersymmetric Localization}.
\newblock {\em JHEP}, 03:040, 2014.

\bibitem{Hosseini:2016cyf}
Seyed~Morteza Hosseini, Anton Nedelin, and Alberto Zaffaroni.
\newblock {The Cardy limit of the topologically twisted index and black strings
  in AdS$_{5}$}.
\newblock {\em JHEP}, 04:014, 2017.

\bibitem{Hosseini:2019lkt}
Seyed~Morteza Hosseini, Kiril Hristov, and Alberto Zaffaroni.
\newblock {Microstates of rotating AdS$_{5}$ strings}.
\newblock {\em JHEP}, 11:090, 2019.

\bibitem{Hosseini:2020vgl}
Seyed~Morteza Hosseini, Kiril Hristov, Yuji Tachikawa, and Alberto Zaffaroni.
\newblock {Anomalies, Black strings and the charged Cardy formula}.
\newblock {\em JHEP}, 09:167, 2020.

\bibitem{Gaiotto:2008ak}
Davide Gaiotto and Edward Witten.
\newblock {S-Duality of Boundary Conditions In N=4 Super Yang-Mills Theory}.
\newblock {\em Adv. Theor. Math. Phys.}, 13(3):721--896, 2009.

\bibitem{Alday:2012au}
Luis~F. Alday, Martin Fluder, and James Sparks.
\newblock {The Large N limit of M2-branes on Lens spaces}.
\newblock {\em JHEP}, 10:057, 2012.

\bibitem{Imamura:2012rq}
Yosuke Imamura and Daisuke Yokoyama.
\newblock {$S^3\mathbb{/}Z_n$ partition function and dualities}.
\newblock {\em JHEP}, 11:122, 2012.

\bibitem{Imamura:2013qxa}
Yosuke Imamura, Hiroki Matsuno, and Daisuke Yokoyama.
\newblock {Factorization of the $S^3/\mathbb{Z}_n$ partition function}.
\newblock {\em Phys. Rev. D}, 89(8):085003, 2014.

\bibitem{Inglese:2023wky}
Matteo Inglese, Dario Martelli, and Antonio Pittelli.
\newblock {The Spindle Index from Localization}.
\newblock 3 2023.

\bibitem{Drukker:2012sr}
Nadav Drukker, Takuya Okuda, and Filippo Passerini.
\newblock {Exact results for vortex loop operators in 3d supersymmetric
  theories}.
\newblock {\em JHEP}, 07:137, 2014.

\bibitem{Assel:2016pgi}
Benjamin Assel, Dario Martelli, Sameer Murthy, and Daisuke Yokoyama.
\newblock {Localization of supersymmetric field theories on non-compact
  hyperbolic three-manifolds}.
\newblock {\em JHEP}, 03:095, 2017.

\bibitem{Cabo-Bizet:2018ehj}
Alejandro Cabo-Bizet, Davide Cassani, Dario Martelli, and Sameer Murthy.
\newblock {Microscopic origin of the Bekenstein-Hawking entropy of
  supersymmetric AdS$_{5}$ black holes}.
\newblock {\em JHEP}, 10:062, 2019.

\bibitem{FRIEDMAN2004362}
Eduardo Friedman and Simon Ruijsenaars.
\newblock Shintani–barnes zeta and gamma functions.
\newblock {\em Advances in Mathematics}, 187(2):362--395, 2004.

\end{thebibliography}

\end{document}